\documentclass[prl,superscriptaddress,floatfix,twocolumn,aps,showpacs]{revtex4}
\usepackage{amssymb} \usepackage{graphicx} \usepackage{longtable}
\usepackage{amsmath} \usepackage{amsfonts} \usepackage{color}
\usepackage{times} \usepackage{stmaryrd} \usepackage[latin1]{inputenc}
\def\mgap{}
\begin{document}
\title{Probing molecular spin clusters by local measurements}
\author{Filippo Troiani}
\email{troiani.filippo@unimore.it}
\affiliation{
National Research Center S3 c/o Dipartimento di Fisica via G. Campi
213/A, CNR-INFM, 41100, Modena, Italy}
\author{Matteo G. A. Paris}\email{matteo.paris@fisica.unimi.it}
\affiliation{Quantum Technology Lab, Dipartimento di Fisica dell'Universit\`{a} degli Studi di
Milano, I-20133 Milano, Italy}
\begin{abstract}
We address the characterization of molecular nanomagnets at the quantum
level and analyze the performance of local measurements in estimating 
the physical parameters in their spin Hamiltonians. To this aim, we 
compute key quantities in quantum estimation theory, such as the classical 
and the quantum Fisher information, in the prototypical case of an 
heterometallic antiferromagnetic ring. We show that local measurements, 
performed only on a portion of the molecule, allow a precise estimate of the 
parameters related to both magnetic defects and avoided level crossings.
\end{abstract}
\date{\today }
\pacs{75.50.Xx, 71.10.Jm, 03.65.Ta}
\maketitle
{\em Introduction}---
Molecular nanomagnets are low-dimensional spin systems, displaying
a variety of nonclassical features \cite{Wernsdorfer,Ardavan,Joris,Bertaina,Friedman}.
The magnetic properties of these systems can be interpreted in
terms of their spin Hamiltonians, which typically depend on a number of
{\mgap unknown coupling constants} \cite{Gatteschi07,Furrer}.
The number of independent parameters can be reduced on the basis
of symmetry arguments, and their values can in principle be computed
from first-principles \cite{Canali,Bellini}. 
However, these approaches are computationally demanding and are affected 
by their own uncertainties. Therefore, the parameters entering the spin 
Hamiltonians are generally obtained by fitting experimental curves \cite{Schnack,Carretta}. 
In particular, when experiments are performed at temperatures lower than 
the energy gap between ground and 
first-excited states \cite{Lascialfari}, 
the estimation of the physical parameters is made 
possible by the dependence on such quantities of the system ground state. 
In fact, any variation in some parameter of interest $\lambda$ modifies 
the ground state, 
and thus the statistics related to the accessible physical observables.  
Any bound to the precision in the estimation procedure 
should thus be connected to the distance between ground states corresponding 
to infinitesimally close values of $\lambda$ \cite{pau06,coz07,zan07}.
Such intuition can be made more rigorous and quantitative upon employing
tools from quantum estimation theory
\cite{lqe1,lqe2,lqe3,lqe5,lqe6}. This allows one to design
optimal estimation procedures and to compute the fundamental limits to
precision, as dictated by quantum mechanics. Indeed, the infinitesimal
(Bures) distance between ground states corresponding to neighboring values 
of $\lambda$ is proportional to the maximum precision in the 
estimation of such parameter, achievable by any possible measurement. 
The connection between the metric structure of the Hilbert space and quantum
estimation theory has in fact been exploited to characterize several
system of interest in quantum technology and to 
address quantum critical systems as a resource for 
quantum estimation \cite{zan,tsa,man,lke}.
\par
In this Letter, we make use of two key quantities in quantum
estimation theory, in order to assess the precision in the estimation of physical 
parameters entering the spin Hamiltonian of molecular nanomagnets. 
These quantities are the classical and the quantum Fisher information 
(FI and QFI, respectively). 
The FI provides, through the Cramer-Rao inequality \cite{Cra46}, 
a lower bound for the uncertainty in the parameter estimation, based on the 
statistics of a given observable.
The QFI gives an upper bound to the FI of any measurement, and thus the best 
possible precision in the estimation allowed by quantum mechanics, for a given 
parametric dependence of the system (ground) state. As a matter of fact, quantum 
estimation theory also provides 
tools to identify the optimal observable, i.e. the observable whose FI 
equals the QFI, thus paving the way for possible practical implementations.
\par
Here we address the characterization of molecular nanomagnets at the quantum
level and analyze the performances of local measurements, realized by addressing 
a portion of the entire compound, as opposed to global ones, requiring access 
to the molecule as a whole.
Our results clearly indicate that fluctuations induced by the total-spin and 
magnetization tunneling at a level anticrossing, or by the introduction of a 
magnetic defect, can be monitored locally, with nearly the ultimate precision 
allowed by quantum mechanics.  
\par
{\em Quantum estimation theory}---We consider a 
spin Hamiltonian $\mathcal{H}$, which depends on an unknown parameter $\lambda$. The value 
of $\lambda$ has to be inferred by performing quantum-limited measurements on the system 
ground state $ |\psi_\lambda\rangle $, and by suitably processing the sample of 
experimental data. The inferred value of the unknown parameter can thus be expressed as 
a function of such data, known as the {\it estimator}, and typically denoted with 
$\hat \lambda$. This is said to be {\it unbiased} if its expectation value coincides with the 
actual value of the parameter $\lambda$.  
The fundamental limit to the precision that can be achieved in the estimate of $\lambda$ is 
given by the quantum Cramer-Rao bound:
$1 / {\rm Var} (\hat\lambda) \le H(\lambda) $, 
where
$H(\lambda)$ is the quantum Fisher information and
${\rm Var} (\hat\lambda)$ is the variance of any unbiased estimator, corresponding 
to the average square distance between $\lambda$ and $\hat\lambda$. 
For a pure state, the QFI is given by 
$H ( \lambda ) = 4 \left[ \langle \partial_\lambda
\psi_\lambda |\partial_\lambda\psi_\lambda \rangle
+ | \langle \partial_\lambda \psi_\lambda 
|\psi_\lambda \rangle |^2 \right]$.
If the ground state is expanded in a parameter-independent basis, the above
derivative reads:
$ |\partial_\lambda \psi_\lambda\rangle = \sum_k (\partial_\lambda c_k )\, 
|k\rangle $, with 
$ c_k ( \lambda ) = \langle k | \psi_\lambda \rangle $.
\par
\begin{figure}[h!]
\begin{center}
\includegraphics[width=0.5\textwidth]{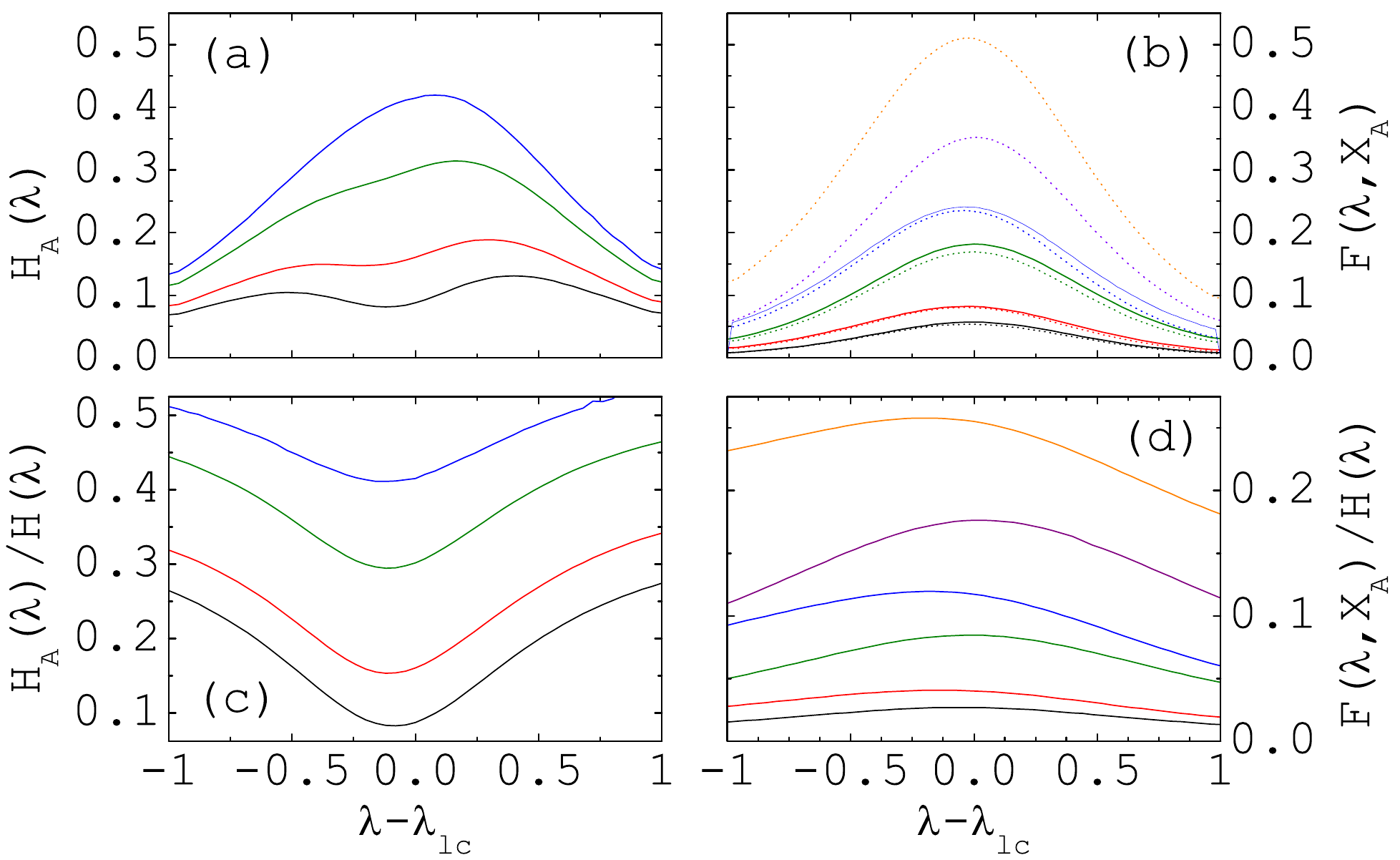}
\caption{\label{Fig1}
(Color online) Quantum estimation of the Cr$_7$Ni molecule ground state at the anticrossing.
We show results for measurements performed on different subsystems $A$ of the ring 
($\alpha/\Delta=1$), formed by the 
first $n_A$ consecutive spins, with $n_A=2$ (black curves), 3 (red), 4 (green), 
5 (blue), 6 (purple), and 7 (orange) respectively. The four panels show:
(a) the QFI; (b) the FI corresponding to the observable
$ X_A \equiv \rho^A_{11} - \rho^A_{22} $ (dotted lines), and QFI obtained for a mixture 
of the diabatic states (solid lines); the (c) QFI and (d) FI of the subsystems, 
normalized to the QFI of the whole ground state. The dotted lines in (b) 
represent the QFI corresponding to the mixture, rather than the linear 
superposition, of the states $|1\rangle$ and $|2\rangle$.}
\end{center}
\end{figure}
\par
If only a specific observable $X$ is available, then the precision of the parameter estimation
is bounded by the classical Cramer-Rao inequality:
$1/{\rm Var} (\hat\lambda) \le F(\lambda , X)$.
Here 
$ F(\lambda , X ) = \sum_x p_\lambda (x) [\partial_\lambda\ln p_\lambda (x)]^2$
is the Fisher information,
and 
$ p_\lambda(x) = |\langle x | \psi_\lambda\rangle|^2 $
is the probability of obtaining the outcome $x$ from the measurement of $X$, 
at a given $\lambda$.
The quantum Cramer-Rao theorem states that 
the FI is bounded from above by the QFI: 
$ F(\lambda , X) \le H(\lambda) $. Any observable $X$ which saturates
the above inequality is said to be {\em optimal}, in that it maximizes
the precision in the estimate of $\lambda$. 
\par
The optimal measurement generally involves accessing the system ground state 
as a whole. A question arises on whether, and to which extent, its performances 
may be emulated by measurements that are local in nature, i.e. performed only
on a portion of the entire system.  Such question can be answered by
evaluating the QFI for the reduced density operator describing a
specific subsystem $A$, as obtained by performing
a partial trace on the complementary subsystem $B$, 
$ \rho_{\lambda}^A = {\rm Tr}_{B}
\left[|\psi_\lambda\rangle\langle\psi_\lambda|\right] $. 
The local QFI is given by the expression 
$ H_A (\lambda) = 2 \sum_{i,j} | \langle \phi_i | \partial_\lambda \rho^A_\lambda | \phi_j \rangle |^2 / (p_i+p_j) $.
Here, $p_i$ and $|\phi_i\rangle$ are the eigenvalues and eigenstates of
$\rho^A_\lambda$, respectively, and the sum is extended over all the indices 
such that $p_i+p_j > 0$.


The above quantities allow a thorough characterization of the parameter
estimation performed through measurements on the system ground state.
In fact, the ratio between FI and QFI quantifies the relative
suitability of the observable $X$ to estimate the parameter $\lambda$.
The ratio $ H_A / H $, instead, assesses to which extent a precise estimate
of $\lambda$ can be obtained by means of local measurements within a
given subsystem $A$.
\par
{\em Level anticrossings}---
In analyzing the ground-state dependence on a physical parameter, a
special attention should be devoted to the avoided level crossings.
Here,  small variations of a physical parameter can induce large changes 
in the system ground state, which are reflected in pronounced 
peaks of the QFI and, possibly, of the FI of some accessible observable. 
Level anticrossings thus represent a resource for the characterization of spin 
Hamiltonians.
For the sake of the following discussion, we write the spin Hamiltonian in 
the generic form 
$ \mathcal{H} = \mathcal{H}_0 + \lambda \mathcal{H}_1 + \mathcal{H}_2$,
where the two dominant terms $ \mathcal{H}_0 $ and $ \mathcal{H}_1 $ 
commute with each other, but not with the small term $\mathcal{H}_2$. 
By varying the parameter $\lambda$ in the vicinity of some critical value
$\lambda_{lc}$, one can induce a level crossing between two joint eigenstates 
of $ \mathcal{H}_0 $ and $ \mathcal{H}_1 $, hereafter denoted by $|1\rangle$ 
and $|2\rangle$.
If these two states are energetically far from all the others for 
$\lambda \simeq \lambda_{lc}$, the system Hamiltonian can be 
effectively reduced to
$ h = \left[\alpha (\lambda-\lambda_{lc}) \sigma_3 + \Delta
\sigma_1\right] / 2$, 
where $\alpha$ is the rate with which the diagonal gap varies as a function of $\lambda$,
$\sigma_j$, $j=1,3$ are Pauli matrices in the basis $\{ |1\rangle , |2\rangle \}$, 
and $ \Delta = 2\langle 1 |\mathcal{H}_2 | 2 \rangle $ is assumed to be 
real and positive. 
The ground state of such effective two-level system can be written as  
$ |\psi_\lambda\rangle = c_1 (y) |1\rangle + c_2 (y) |2\rangle $, where 
$ y \equiv \alpha (\lambda-\lambda_{lc})/ \Delta$ represents the (normalized) 
distance of the parameter $\lambda$ from the critical value $\lambda_{lc}$.
It follows that the FI corresponding to a generic observable $X$ can be written 
in the product form $ F = (\alpha / \Delta)^2 f_X(y)$, where the function $f_X$ 
is given by the following expression:
\begin{equation}\label{eqFI}
f_X\!\!=\! \frac{y\!+\!\sqrt{1+y^2}}{2(1+y^2)^{5/2}}\!\sum_x\!
\left\{\!\!
\frac{\langle 1 | x \rangle^2\!-\!\langle 2 | x \rangle^2\!+\!2y \langle 1 | 
x \rangle \langle 2 | x \rangle}{[y+\sqrt{1+y^2}]\langle 1 | x \rangle -\langle 2 | x \rangle} 
\!\right\}^2 .
\end{equation}
As detailed in the Supplemental Material \cite{supp}, also the quantum Fisher 
information can be written in a factorized form:  
\begin{equation}\label{eqQFI}
H 
= 
 \frac{(\alpha/\Delta)^2}{\{1+[\alpha(\lambda-\lambda_{lc})/\Delta]^2\}^{2}} 
\equiv 
(\alpha / \Delta)^2 f_H (y) .
\end{equation}
The above functions $f_X$ and $f_H$ thus                                                                                                                                      specify the dependence of the highest precision achievable in the parameter estimation on the 
distance $y$ from the crossing point. The presence of the prefactor 
$(\alpha /\Delta)^2$ quantifies the increase of the precision that can be achieved, 
for each given distance $y$, by making the anticrossing narrower.  
Besides, from Eq. (\ref{eqFI}) it follows that an observable $X$ is 
optimal if there are two measurement outcomes $x$ and $x'$ allowing for 
a perfect discrimination between any two orthogonal states, i.e. if $ | x \rangle $ 
and $ | x' \rangle $ are orthogonal linear superpositions of the $|1\rangle $ and 
$ |2\rangle$ states. 
In this case, in fact,
$f_X (y) = f_H (y) $, and the FI of $X$ equals the QFI. It can be easily verified that 
(in the absence of degeneracy at the distance $y$ of interest) both $\mathcal{H}_0$ 
and $\mathcal{H}_1$ fulfill the above condition, and thus represent optimal observables.
\par
\begin{figure}[h!]
\begin{center}
\includegraphics[width=0.5\textwidth]{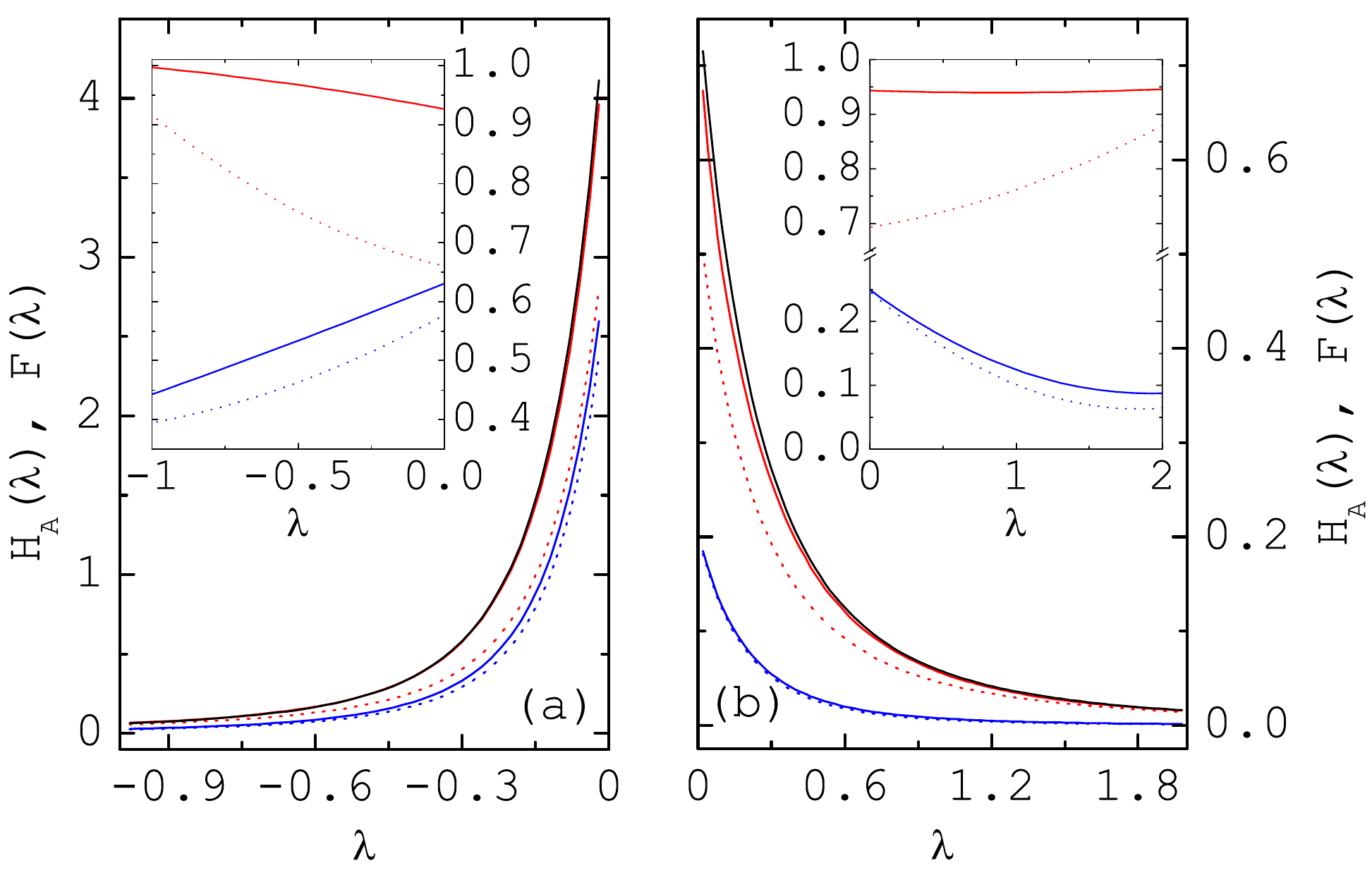}
\caption{\label{Fig2}
(Color online) 
Quantum estimation of the exchange coupling between the magnetic defect and 
the neighboring spins (Cr-Ni), with respect to that between all the other 
neighboring spins (Cr-Cr). QFI (solid curves) and FI (dotted) corresponding to 
two- (blue) and 
three-spin (red) subsystems. Panels (a) and (b) correspond to ferromagnetic and 
antiferromagnetic Cr-Ni coupling, respectively. In the insets of the two panels, the
same quantities are normalized to the QFI of the whole ground state.}
\end{center}
\end{figure}
\par
{\em Numerical results}---The problem of estimating the physical parameters 
that enter the spin Hamiltonian is ubiquitous in molecular magnetism. In the
following, we consider in some detail the representative example of the 
Cr$_7$Ni molecule. Its magnetic core is formed by seven 
Cr$^{3+}$ ions, each carrying an $s_{\rm Cr}=3/2$ spin, and one Ni$^{2+}$ ion, 
with $s_{\rm Ni}=1$ \cite{troiani}. 
As a spin ring with dominant antiferromagnetic exchange interaction, Cr$_7$Ni 
represents a prototypical model of a highly-correlated, low-dimensional quantum 
system \cite{siloi}. Besides, the presence in such molecule of the Ni ion allows 
us to extend the present discussion to the role of magnetic defects. 
Given the purpose of the present paper, we focus on the functional dependence 
of the FI and QFI on the main physical parameters entering the spin Hamiltonian, 
rather than on their specific values, as estimated by different experimental and 
theoretical means. 
\par
As an example of an anticrossing in the system ground state, we consider the 
one between the lowest eigenstates of $ \mathcal{H}_0 + \mathcal{H}_1 $ with 
$S=M=1/2$ and $S=M=3/2$, hereafter labeled $|1\rangle$ and $|2\rangle$, respectively.
The two terms of the Hamiltonian involved in the level crossing account  
for the exchange interaction between neighboring spins,
$
\mathcal{H}_0 = J \sum_{k=1}^{8} {\bf s}_k \cdot {\bf s}_{k+1} 
$ (with $J>0$), 
and for the coupling to an applied magnetic field,
$ \mathcal{H}_1 = - \lambda \alpha S_z $.
The unknown parameter $\lambda$ thus coincides with the Zeeman splitting 
in units of $\alpha$, and can be identified for example with the $g$-factor of the 
molecule for $ \alpha = \mu_B B$. The zero-field gap between the ground $S=1/2$ 
doublet and the lowest $S=3/2$ quadruplet, mainly induced by the exchange interaction, 
determines the value of $\lambda_{lc}$. 
The small term $ \mathcal{H}_2 $ includes all the remaining contributions in 
the spin Hamiltonian, which are responsible for the gap $\Delta$ \cite{troiani}.
\par
The dependence on $\lambda$ of the system ground state and of the 
corresponding reduced density operators is summarized by the behavior
of the FI and of the QFI. In particular, 
three main features emerge from the $H_A(\lambda)$. First, for 
subsystems $A$ formed by a small number of consecutive
spins ($n_A=2,3,4$), the highest values of $H_A$ are obtained away from the 
crossing point, where the QFI presents instead a clear dip [see Fig. \ref{Fig1}(a)]. 
Second, such feature can be linked to the phase coherence between the 
states $|1\rangle$ and $|2\rangle$ that contribute to the ground 
state $ |\psi_\lambda\rangle $. In fact, the value of the QFI corresponding 
to the mixture 
$ \sigma_{\lambda}^A = c_1^2(y) \rho^A_{11} + c_2^2(y) \rho^A_{22} $ 
presents lower values for all $\lambda$s, and a maximum close to $ \lambda = 
\lambda_{lc} $ [solid lines in Fig. \ref{Fig1}(b)]. 
The QFI of $ \sigma_{\lambda}^A $ also corresponds to the maximum of the
 FI of $|\psi_\lambda\rangle$, restricted to observables $X$ that are diagonal 
 in the basis of the diabatic states $\{ |1\rangle , |2\rangle \}$ \cite{supp}.
Therefore, the comparison between the QFI of $\rho^A_\lambda$ and $\sigma^A_\lambda$
shows that the performance of a local observable $X$ at an avoided level crossing 
can in general benefit from the fact that $X$ is not diagonal in the basis of the diabatic states. 
Third, within these observables, the operator $ X_A \equiv \rho^A_{11} - \rho^A_{22} $, 
with $ \rho_{ij}^A = \hbox{Tr}_B (|i \rangle\langle j|) $,
is approximately optimal (dotted lines). 
Finally, we note that not only the maximum of the QFI of local observables can be localized 
away from the crossing point, but $\lambda_{lc}$ also corresponds to an absolute minimum
 for the relative suitability of the local measurements.
This clearly emerges from the plots of $H_A(\lambda)$ and $F(\lambda , X_A)$, normalized 
to the QFI of the ground state [Fig. \ref{Fig1}(c,d)].
\par
In order to gain some quantitative insight into the problem, we consider the case where 
the actual value of the unknown parameter $\lambda = g $ is 2, and this coincides with 
the critical value, given the applied magnetic field $B$.  
In this case, the mean squared error in the estimate of the $g$-factor resulting from a 
single quantum measurement is given by
Var$^{1/2}(\hat\lambda) = (\Delta / \mu_B B) [H_A (y=0)]^{-1/2}$, 
which for two spins (black curves), is approximately $0.05$
(we have taken $B=10\,$T, which approximately corresponds to the field that induces 
the level crossing between the $S=M=1/2$ and the $S=M=3/2$ eigenstates, and 
$\Delta = 0.1\,$K, which is a typical value of the gap in the Cr-based rings).
The mean squared error can in principle be reduced by a factor $\sqrt{N}$ by passing 
from a single measurement to a set of $N$ measurements, or by working at a narrower 
anticrossing. 
\par
\begin{figure}[h!]
\includegraphics[width=0.5\textwidth]{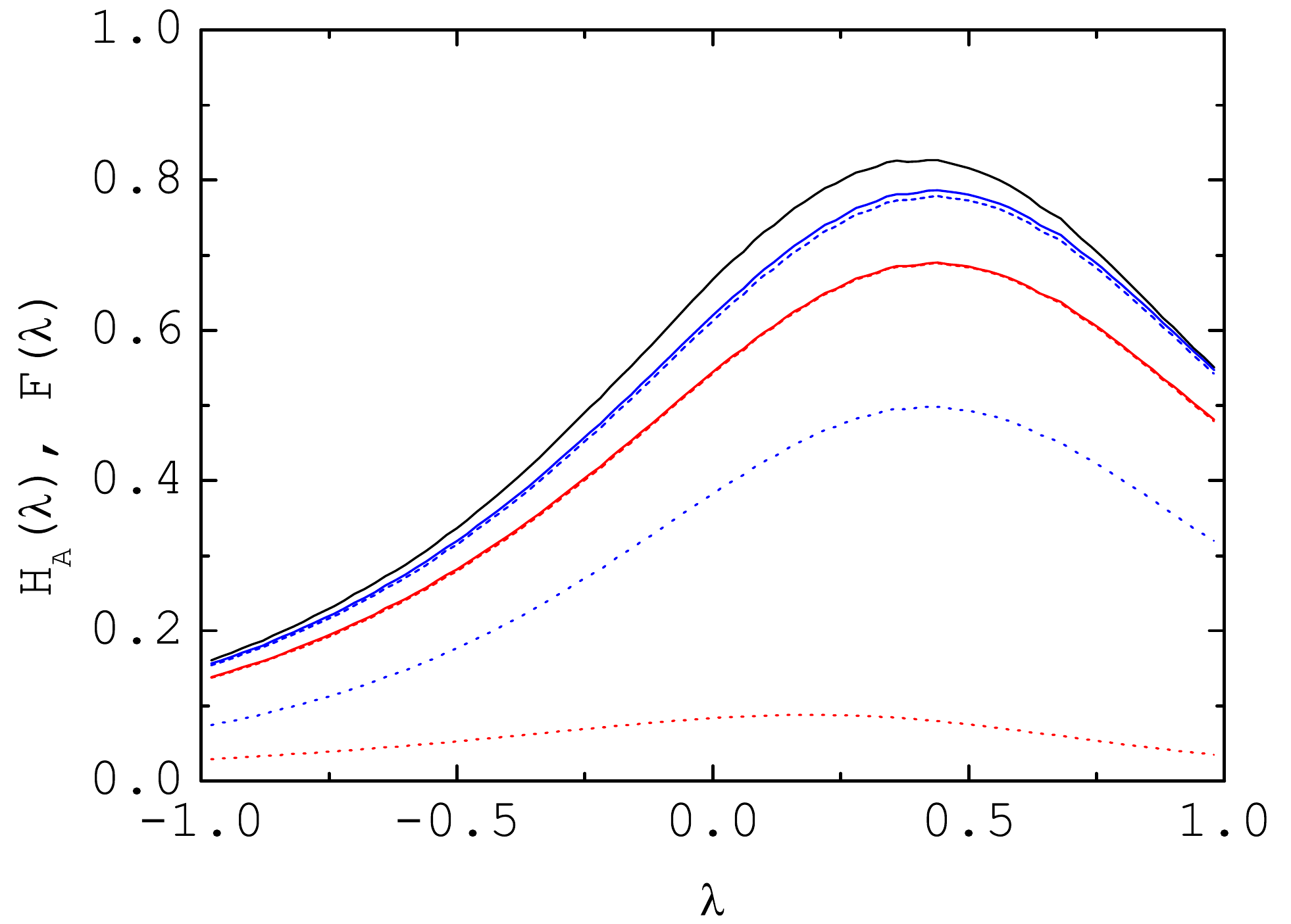}
\caption{\label{Fig3}
(Color online) 
Quantum estimation of the inhomogeneity in the $g$-factor associated with the 
magnetic defect, i.e. $\mu_B(g_{\rm Ni}-g_{\rm Cr})B/J$. 
Solid curves correspond to the QFI: for the whole system (black), for subsystems 
formed by two (red) or three spins (blue).
Dashed and dotted curves correspond to the FI for the local magnetization, with 
and without resolution between the spins of the subsystem.}
\end{figure}
{\em Exchange interaction}---We next consider a ground state that changes 
gradually with $\lambda$, away from a level crossing 
(here, $\mathcal{H}_0$ and $\mathcal{H}_1$ don't commute, and $\mathcal{H}_2$ 
can be set to zero). In particular, we are interested in the case where the 
unknown parameter is related to a magnetic defect, such as the $s_8=s_{\rm Ni}$
spin in the Cr$_7$Ni molecule. This spin represents a defect because its length 
differs from that of all the other spins in the ring. Besides, the Cr-Ni 
exchange coupling can differ from the Cr-Cr ones, and the Ni $g$-factor can 
differ from that of the Cr ions. We start by considering the effect of an 
inhomogeneous exchange interaction, and correspondingly group the relevant part 
of the spin Hamiltonian into the two terms: 
\begin{equation}\label{eqH1}
\mathcal{H}_0 = J \sum_{k=1}^{6} {\bf s}_k \cdot {\bf s}_{k+1}, \  
\mathcal{H}_1 = \lambda J {\bf s}_8 \cdot ( {\bf s}_{7} + {\bf s}_{1}),
\end{equation}
where the unknown parameter $\lambda$ coincides with the ratio between the Cr-Ni
and Cr-Cr exchange couplings.
\par
The dependence of the system ground state on $\lambda $ is characterized in 
terms of the QFI $H(\lambda)$ (Fig. \ref{Fig2}), both for negative and positive
values of the parameter [panels (a) and (b), respectively]. For $ \lambda < 0 $,
the defect is ferromagnetically coupled to its neighbors, and the system ground 
state has $S=5/2$. For $ \lambda > 0 $, instead, such coupling is antiferromagnetic, 
and the total spin is $S=1/2$. 
In both cases, $H(\lambda)$ is maximal for $ \lambda \rightarrow 0 $, 
and decreases monotonically with $|\lambda|$ (solid black curves). 
The distinguishability between 
two (infinitesimally) close values of $\lambda$ is thus relatively large in the 
weak-coupling limit, while the ground state is weakly dependent on the precise 
value of $ \lambda $ in the (more realistic) range of values $\lambda \simeq 1$.
In the considered range of parameters, the lowest mean squared error that can be 
achieved in the estimate of the Cr-Ni exchange coupling by means of a single 
quantum measurement, Var$^{1/2}(\hat\lambda) = J [H (\lambda)]^{-1/2} $, is of 
the order of the Cr-Cr exchange coupling $J$.
Besides the absolute value of the QFI, we are interested here in the comparison 
between the QFI corresponding to the ground state and the same quantity derived 
for the reduced density operators. We note that, already for subspaces $A$ 
formed by three consecutive spins (solid red), $ H_A (\lambda) $ approaches 
$ H (\lambda) $. The QFI corresponding to two-spin subsystem (solid blue), 
instead, approaches $H(\lambda)$ only for $ \lambda < 0$. The ratios between the 
local QFI and that of the whole ground state are reported in the figure insets.
Therefore, local observables are in principle well suited for precisely estimating 
the exchange coupling between the magnetic defect and the neighboring spins.
Interestingly, local observables consisting of exchange operators,
$
X (n_A) = \sum_{i=k}^{k+n_A-1} {\bf s}_i \cdot {\bf s}_{i+1} 
$,
are nearly optimal.
This is shown by the FI corresponding to $n_A=2$ and $n_A=3$ (dotted curves), 
which are very close to the QFI of the corresponding subsystems. The Fisher 
information of the local magnetization (not shown) gives instead 
significantly lower values. 
\par
\par{\em Magnetic field}---The magnetic defect affecting the ground state 
can also consist in the presence of a spin with a different $g$-factor 
(or, equivalently, in a local magnetic field). 
In this case, the relevant terms of 
the spin Hamiltonian are grouped as follows:
\begin{equation}
\mathcal{H}_0 = J \sum_{k=1}^{8} {\bf s}_k \cdot {\bf s}_{k+1} 
, \  \mathcal{H}_1 = \lambda J s_{N,z} .  
\end{equation}
The unknown parameter $\lambda$ thus corresponds to the difference in the 
Zeeman splitting of the Ni ion with respect to that of the Cr ions,
normalized to the exchange coupling, 
$\lambda = \mu_B(g_{\rm Ni}-g_{\rm Cr})B/J$.
The quantum Fisher information of the system ground state
(solid black line in Fig. \ref{Fig3}) presents a pronounced maximum for 
$\lambda \simeq 0.5 $. As in the previous case, the QFI information 
corresponding to two-and three-spin subsystems (solid blue and red, 
respectively) approaches $H(\lambda)$, especially if the subsystems $A$
includes the defect. 
The FI corresponding to the local magnetization, 
$ 
X(n_A) = \sum_{i=k}^{k+n_A} a_i s_{i,z} 
$,
falls significantly below the QFI for the corresponding subsystem if the 
observable is not spin selective ($a_i = a_j$ for all $i \neq j$, dotted lines). 
However, if the magnetization is spin selective ($a_i \neq a_j$ for $i \neq j$, 
dashed lines), 
the values of the FI are very close to the maximal ones. In the latter case, the 
magnetization thus represents a nearly optimal observable for the parameter 
estimation.
\par
{\em Conclusions}---We have analyzed the performances of local measurements 
in estimating different physical parameters that enter the spin Hamiltonian of a 
molecular nanomagnet. Local measurements are shown to allow a precise estimation 
of parameters related to both magnetic defects and avoided level crossings. Parameters 
such as the exchange coupling or the $g$-factor of a magnetic defect can be estimated 
locally ---with nearly the ultimate precision allowed by quantum mechanics--- by measuring 
related observables, namely the exchange operators and the local magnetization, 
respectively. Local measurements also approach the ultimate precision in the parameter 
estimation at avoided level crossings, where the commutation relations between the 
observable and the Hamiltonian are shown to play a relevant role. Our results 
clearly show the effectiveness of local measurements in probing Hamiltonian parameters, 
thus paving the way for the development of optimal characterization schemes for 
molecular spin clusters. 
\par
This work was funded by: the Italian Ministry of Education and Research, through the FIRB project RBFR12RPD1 and the Progetto Premiale EOS; the US AFOSR/AOARD program (contract FA2386-13-1-4029); the EU, through the Collaborative Projects QuProCS (Grant Agreement 641277) and the FP7 FET project $MoQuaS$ (contract N.610449); the UniMI, through the H2020 Transition (Grant 14-6-3008000-625).

\section*{SUPPLEMENTAL MATERIAL}

\subsection{Derivation of the functions $f_H$ and $f_X$}

The ground state of the effective Hamiltonian $h$ can be expressed as a function of the basis states $|1\rangle$ and $|2\rangle$ by means of the coefficients
\begin{equation}
c_1(y) = P(y) Q(y) , \ c_2(y) = - Q(y) ,
\end{equation}
where 
$ y = \alpha (\lambda - \lambda_{lc}) / \Delta $
is the normalized distance of the parameter $\lambda$ from the critical value.
The functions $P$ and $Q$ are given by the following expressions:
\begin{equation}
P(y) = y + \sqrt{1+y^2} , \ Q^{-1}(y) = \sqrt{2 P(y)} \, (1+y^2)^{1/4}.
\end{equation}
From the above equations, it follows that the derivatives of the coefficients, 
entering the expressions of both the classical and the quantum Fisher information, 
are given by:
\begin{equation}
\partial_y c_1(y) = \frac{Q(y)}{2(1+y^2)} , \ 
\partial_y c_2(y) = \frac{P(y)\,Q(y)}{2(1+y^2)} .
\end{equation}
As a result, the expression of $H(\lambda)$ takes the form:
\begin{equation}
H = 4 (\alpha/\Delta)^2 \left[ (\partial_y c_1)^2 + (\partial_y c_2)^2 \right]
= \frac{(\alpha/\Delta)^2}{(1+y^2)^2} ,
\end{equation}
where we made use of the equation
$ \partial_\lambda = (\alpha/\Delta) \partial_y $.
As to the classical Fisher information corresponding to the observable $X$, 
this can be written as a function of the amplitudes $ \langle 1 | x \rangle $
and $ \langle 2 | x \rangle $ (which are assumed to be real, for simplicity)
and of their derivatives with respect to $\lambda$ (or $y$). 
These enter the expression of the probabilities 
\begin{equation}
p_\lambda (x) = \langle \psi_\lambda | x \rangle^2 = 
\sum_{k,l=1}^2 c_k(y) c_l(y) \langle k | x \rangle \langle x | l \rangle .
\end{equation}
The derivative of such probability with respect to $y$ can be shown to be:
\begin{equation}
\partial_y p_\lambda (x) = 
\frac{P(y)[Q(y)]^2}{1+y^2} 
(\langle 1 | x \rangle^2 \!\!-\!\! \langle 2 | x \rangle^2 + 
2 y \langle 1 | x \rangle\langle 2 | x \rangle).
\end{equation}
After replacing the two above expressions into that of the Fisher information, 
\begin{equation}
F ( \lambda , X ) = (\alpha/\Delta)^2 \sum_x \frac{[\partial_y p_\lambda (x)]^2}{p_\lambda (x)},
\end{equation}
one can derive the Eq. (1) reported in the manuscript.
\par
In order to highlight the role of the phase coherence between the two basis states, the QFI of $|\psi_\lambda\rangle$ can be compared with that obtained for the statistical mixture of $|1\rangle$ and $|2\rangle$, with populations corresponding to $[c_k(y)]^2$.
In this case, the probabilities $p_\lambda (x)$ take the form
\begin{equation}
p^{inc}_\lambda (x) =  
\sum_{k=1}^2 [ c_k(y) \langle k | x \rangle ]^2 .
\end{equation}
The corresponding derivative with respect to $y$ reads
\begin{equation}
\partial_y p^{inc}_\lambda (x) = 
\frac{P(y)[Q(y)]^2}{1+y^2} 
(\langle 1 | x \rangle^2 \!\!-\!\! \langle 2 | x \rangle^2) .
\end{equation}
The adimensional function that enters the expression of the quantum Fisher information thus becomes:
\begin{equation}
f_X^{inc}\!\!=\! \frac{y\!+\!\sqrt{1+y^2}}{2(1+y^2)^{5/2}}\!\sum_x\!
\frac{(\langle 1 | x \rangle^2\!-\!\langle 2 | x \rangle^2)^2}
{[(y+\sqrt{1+y^2})\langle 1 | x \rangle]^2 + \langle 2 | x \rangle^2} .
\end{equation}
This also corresponds to the function $f_X$ for an observable 
$ X = \sum_x x |x\rangle\langle x| $, 
which is diagonal in the basis of the diabatic states, and thus such that 
$\langle 1 | x \rangle\langle x | 2 \rangle = 0$ for any $x$. This follows simply 
from the fact that, for such an observable, $p_\lambda (x) = p^{inc}_\lambda (x)$.
\par
We consider the case where there are two outcomes of the measurement of $X$, $x$ and $x'$, with corresponding eigenstates $|x\rangle$ and $|x'\rangle$ that are mutually orthogonal. We write them as linear combinations of the basis states, with real coefficients (what follows can be easily generalized to the case of complex coefficients): 
$ |x \rangle = a |1\rangle + b |2\rangle $
and 
$ |x'\rangle = b |1\rangle - a |2\rangle $. 
Plugging these expressions into the Eq. (1) of the manuscript, one obtains, after some algebra, the equation 
$f_X = f_H = 1 / (1+y^2)^2$, which implies that the measurement is optimal.
\par
In the case of the Cr$_7$Ni ring, the observable $X_A \equiv \rho_{11}^A - \rho_{22}^A $ fulfils the above condition. In fact, $|1\rangle$ and $|2\rangle$ are eigenstates of $S_z$, corresponding to different values, $M_1=1/2$ and $M_2=3/2$, of the total spin projection. The reduced density operators $\rho_{kk}^A$ (and thus $X_A$) can be written as mixtures of density operators, each with a defined value of the total spin projection. 
This follows from the fact that each finite term of $\rho_{kk}^A$ comes from contributions like $ \langle i_B |k \rangle\langle k| i_B \rangle $, with $|i_B\rangle$ a basis state of the subsystem $B$, which can be chosen so as to have a defined value of the spin projection $M_B$. The ket and the bra in the term of $\rho_{kk}^A$ thus have to be characterized by the same value of $M_A=M_k-M_B$. As a result, $\rho_{kk}^A\otimes I_B$ cannot have matrix elements between states with different values of the total spin projection, such as $|1\rangle$ and $|2\rangle$. 

\subsection{Numerical calculations}

The eigenstates of Cr$_7$Ni are obtained by numerically diagonalizing the Hamiltonian, with the inclusion of the exchange and of the Zeeman terms. The Hamiltonian is computed and diagonalized within the irreducible tensor operator formalism (see, e.g., E. Liviotti (2002) in the references). In the case of the avoided level crossing, the Hamiltonian commutes with ${\bf S}^2$ and $S_z$, and can be diagonalized independently within each $(S,M)$ subspace, with $S=M=1/2$ (dimension 574) and $S=M=3/2$ (dimension 1000). The eigenstates are the expanded in a local basis $|m_1,m_2,\dots ,m_8\rangle$ (with $m_i$ the projection of the $i$-th spin along $z$), and the terms $\rho^A_{ij}$ are computed by performing a partial trace over the spins that don't belong to the subsystem of interest $A$. The reduced density operators $\rho_\lambda$ is then computed by combining the above operators, through the expression
$ \rho^A_\lambda = \sum_{i,j=1}^2 c_i(\lambda)c_j(\lambda) \rho^A_{ij} $. 
This matrix is diagonalized numerically, for all the values $\lambda_k = k \, \delta\lambda$ of the parameter $\lambda$ in the grid, so as to obtain the eigenvalues $p_i$ and the eigenvectors $|\phi_i\rangle$ that enter the expression of $H_A$, for each point of the grid. The derivative of the reduced density operator, $\partial_\lambda \rho_\lambda^A $, is computed numerically as 
$ (\rho^A_{\lambda_{k+1}} - \rho^A_{\lambda_{k-1}}) / (2\delta\lambda) $.
\par
The introduction of the magnetic defect reduces the symmetry of the Hamiltonian. In particular, in the case of the exchange coupling the ground state of the spin Hamiltonian belongs either to the $S=5/2$ or to the $S=1/2$ subspaces, depending on whether the Cr-Ni coupling is ferromagnetic or antiferromagnetic, respectively. In the case of the magnetic field, $\mathcal{H}_1$ doesn't commute with ${\bf S}^2$. This implies that the ground state has to be calculated in a larger subspace, including all the basis states with total spin from $1/2$ to $S_{max}>1/2$. The value of $S_{max}$ is determined upon convergence of the ground state energy and depends on the value of $\lambda$.

\end{document}